\documentclass[aps,prd,preprint,groupedaddress,nofootinbib,floatfix]{revtex4-1}
\renewcommand{\thetable}{\arabic{table}} %%%
\usepackage{amsmath}
\usepackage{graphicx}
\usepackage{amsbsy}
\usepackage{amssymb}
\usepackage{palatino,avant,graphicx,color}
\usepackage[none]{hyphenat}
\usepackage{tikz}
\usepackage{xy}
\usepackage{changes}
\usepackage{wrapfig}
\usepackage{float}
\usepackage{tabularx}
\usepackage{natbib}
\usepackage{hyperref}
\usepackage{mathtools}
\makeatletter
\newcommand{\printfnsymbol}[1]{%
  \textsuperscript{\@fnsymbol{#1}}%
}
\makeatother

\begin{document}

\title{The effect of DNA bases permutation on surface enhanced Raman scattering spectrum}

  \author{Shimon Rubin$^{*}$}
  \author{Phuong H.L. Nguyen$^{*}$}
%\thanks{}
% \email{}
 \author{Yeshaiahu Fainman}
 \affiliation{$^{*}$ Equal contribution}
 \affiliation{Department of Electrical and Computer Engineering, University of California, San Diego, 9500 Gilman Dr., La Jolla, California 92023, USA}
%% \today

\begin{abstract}
Surface enhanced Raman scattering (SERS) process results in a tremendous increase of Raman scattering cross section of molecules adsorbed to plasmonic metals and influenced by numerous physico-chemical factors such as geometry and optical properties of the metal surface, orientation of chemisorbed molecules and chemical environment. While SERS holds promise for single molecule sensitivity and optical sensing of DNA sequences, more detailed understanding of the rich physico-chemical interplay between various factors is needed to enhance predictive power of existing and future SERS-based DNA sensing platforms. In this work we report on experimental results indicating that SERS spectra of adsorbed single-stranded DNA (ssDNA) isomers depend on the order on which individual bases appear in the 3-base long ssDNA due to intra-molecular interaction between DNA bases. Furthermore, we experimentally demonstrate that the effect holds under more general conditions when the molecules don't experience chemical enhancement due to resonant charge transfer effect and also under standard Raman scattering without electromagnetic or chemical enhancements. Our numerical simulations qualitatively support the experimental findings and indicate that base permutation results in modification of both Raman and chemically enhanced Raman spectra. 
 
\end{abstract}

\maketitle

\section*{Introduction}
\sloppy

Surface enhanced Raman scattering (SERS) \cite{fleischmann1974, Jeanmaire1977, albrecht1977} is an intriguing effect which results in several orders amplification of Raman scattering cross section of molecules chemisorbed or physiosorbed to a rough metal surface.
In particular, SERS enhancement factors (EFs), which characterize increase in molecular Raman scattering relative to under non-SERS condition \cite{le2008principles}, mostly stem from conjointly operating 
%concurring 
%and multiplicative 
electromagnetic enhancement (EM) and chemical enhancement (CE) effects.
While EM effect stems from local increase of electromagnetic fields in so-called hot spots which contribute to both larger induced Raman dipole and intensified radiation of the vibrating dipole through excitation of plasmon resonances in the metal nanostructure (operating as plasmonic nanoantenna \cite{moskovitz, stockman}), 
%leading to non-selective enhancement of the relevant Raman active modes; 
CE effect \cite{leru2007,morton2011,leru2013} stems from changes of the Raman polarizability and of the molecular electronic structure due to formation of chemical bonds between the relevant molecule and the metal 
and therefore selectively operates on the formed metal-molecule complex depending on its' specific properties \cite{moskovitz,jensen2008}.
In particular, CE effect  % stiles2008
may be accompanied by either resonant charge transfer between the adsorbed molecule and the metal substrate \cite{kneipp2016} facilitated by formation of charge-transfer states, or non-resonant (static) ground-state charge redistribution due to overlap of metal and molecular orbitals \cite{morton2009} and surface binding (see \cite{ding2017,langer2019} and references within) leading to static far-from-resonance changes of the molecular polarizability.

Due to the high sensitivity of the SERS process inherent to Raman-based vibrational spectroscopy with narrow spectral lines, leading (under some conditions)  to single molecule detection \cite{nie1997,kneipp1997} as well as single base sensitivity \cite{xu2015label,guerrini2015} in simple 
%deoxyribonucleic acid 
DNA sequences, SERS-based DNA sensing holds promise for numerous future bio-related applications such as medical diagnosis, bio-analysis, and environmental monitoring where composition and sequencing analysis of the information-rich DNA molecules serve as one of the central tools. 
Furthermore, the significant progress in more recent years allowing individual nucleobases resolution \cite{he2018} achieved by employing the closely related tip-enhanced Raman spectroscopy (TERS) \cite{stockle2000, anderson2000, hayazawa2000, pettinger2000}, 
%and low-temperature and ultrahigh-vacuum TERS \cite{zhang2013, chiang2015, tallarida2017}  
and exploitation of CE effect \cite{freeman2014} to achieve higher specificity in ssDNA sensing, both indicate potential applications of SERS/TERS for DNA sensing. 
\textcolor{black}{The latter include DNA characterization with a reduced number of thermocycling steps in polymerase chain reaction (PCR) amplification \cite{wu2020performance}, mutation detection due to molecular conformation change \cite{kowalczyk2019} or employing branched DNA technique which relies on the signal amplification of the probe that binds to a specific nucleotide sequence \cite{cheng2018},
which are typically more sensitive than conventional fluorescent labeling (see for instance a recent work \cite{fu2019}).}
Nevertheless, the extreme sensitivity of SERS with respect to numerous factors such as substrate quality, dielectric environment, as well as unknown molecule orientation, occasionally leads to controversial results on gold \cite{harroun2018controversial} and silver \cite{freeman2016} substrates, indicating that further research is needed to understand the basic physical mechanisms of plasmon-DNA interaction and how to leverage them in order to enhance present capabilities, eventually leading to bio-medical applications which require higher precision than achieved today \cite{pyrak2019}. 
In particular, considering increasingly complex DNA sequences, where each nucleotide in turn is composed of additional sub-units, 
%which are phosphate backbone, five-carbon sugar and nitrogenous base, 
is expected to introduce novel mechanisms that affect the corresponding Raman and SERS spectra; 
it is reasonable to assume that incorporating these effects might improve the accuracy of future plasmonic-based detection schemes.

\begin{figure}
	\includegraphics[scale=0.1]{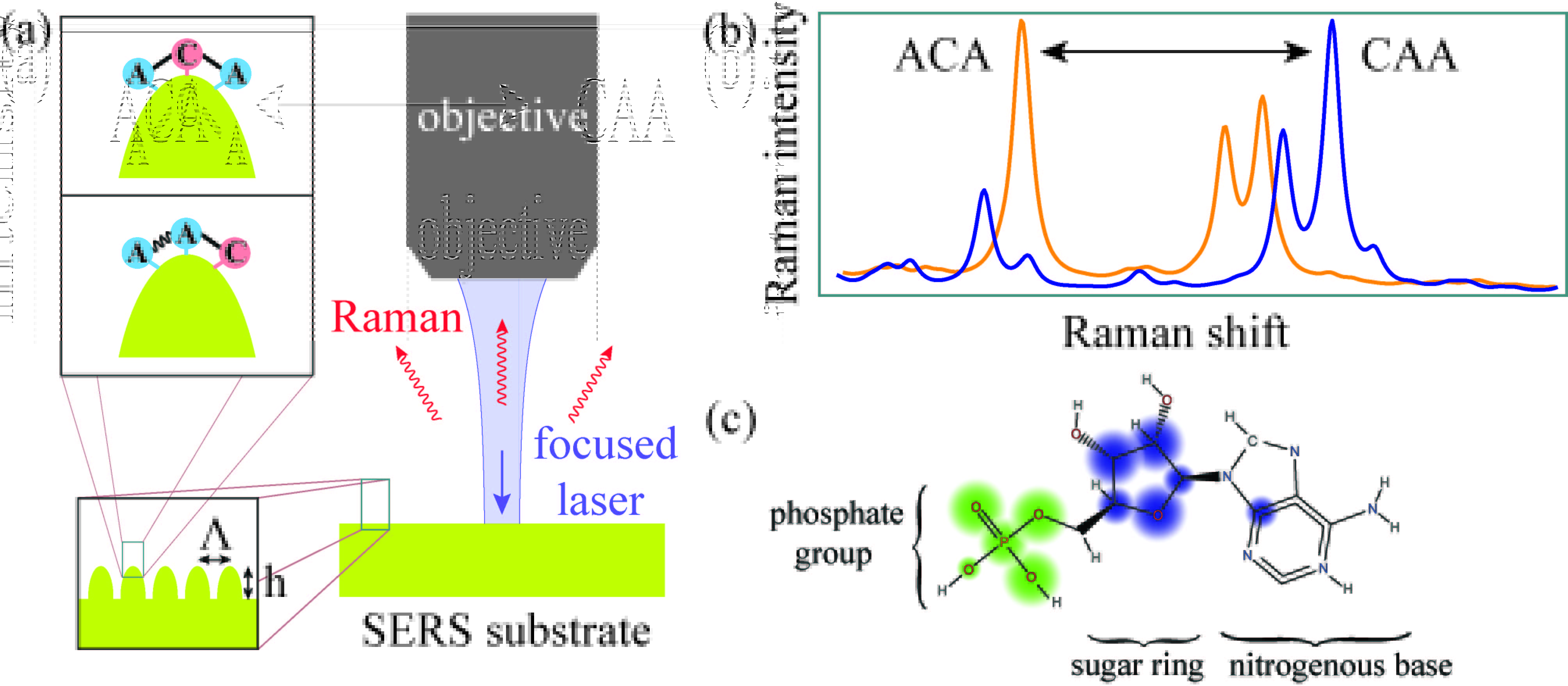}
    \caption{Schematic description of the underlying concept of our work. (a) Key elements of the experimental setup presenting a silver/gold nanorod array substrate of mean period $\Lambda$ and height $h$, hosting ssDNA 3-mers allowing to compare SERS spectra of ACA with CAA and of CAC with CCA (only ACA and CAA are presented in the scheme). The permutation of DNA bases in the sequences leads to a change in the interaction between the DNA bases, schematically represented as different bonds between the bases, and to enhancement of different molecular vibrational modes with distinct dominant frequency (b) and different kinetic energy distribution among the different atom groups (c); relative disks' size indicate the corresponding kinetic energy stored in the relevant atoms at the dominant mode.}
    \label{Scheme}
\end{figure}

In this work we experimentally and numerically study, \textcolor{black}{for the first time to the best of our knowledge,} the effect of intra-molecular interaction between DNA bases on the resultant SERS and Raman spectra, by considering simple ssDNA sequences which admit identical number of bases appearing at different order along the sequence.
Taking advantage of 
the controlled distance between DNA bases in ssDNA molecules allows us to investigate how the position of DNA bases 
and their nearest neighbors affect the frequency, RA (Raman activity) and vibrational pattern of the corresponding normal modes.  
Fig.\ref{Scheme} presents schematic description of the key elements in our study which includes silver/gold nanorod array serving as a substrate for adsorbed ssDNA 3-mer strands, allowing to extract changes of the SERS spectra due to  nucleobases permutation, as well as different units/groups in a single DNA base which are activated depending on their position in ssDNA molecule. 
In particular, we experimentally measure SERS spectra of two groups of isomers classified according to the total number of nucleotides of a given type; group ($1$) comprised of ACA and CAA sequences which admit two adenines (A) and a single cytosine (C), and group ($2$) comprised of CAC and CCA which admit a single A and two C's.
We then analyze the corresponding SERS spectra and compare them by
constructing the corresponding statistical correlation matrices which provides a quantitative metric to characterize the changes of the spectra due to permutation of DNA bases.
In addition to the experiments where ssDNA molecules adsorbed directly to metal nanorod substrate array which facilitates both EM and CE effects, we also consider Raman spectra of ssDNA molecules in bulk solution without CE and EM effects as well as ssDNA molecules adsorbed to gold and silver nanorod substrate arrays covered with a $2$-nm thickness of Al$_2$O$_3$ dielectric film. The latter is thin enough to facilitate EM effect but eliminate the CE effect, allowing to study the difference between the various spectra of ssDNA molecules without the charge transfer resonant effect \cite{freeman2014}. 
Furthermore, we employ density functional theory (DFT) which became a widespread method for calculating electronic excitations and related optical properties of molecules \cite{Ullrich2011} (including Raman polarizability), to study the permutation effect on Raman spectra of ssDNA 3-mers with and without the CE effect. 
The latter is achieved by introducing Ag$_{4}$ silver nano-clusters \cite{huang2013} bond to each one of the nucleotides, allowing to study the effect of static non-resonant (i.e. far from resonance) CE effect on Raman spectra and kinetic energy distribution due to electrostatic charge redistribution. 
For simplicity, our computational model studies single orientation of DNA bases relative to the metal, and therefore it is not used for quantitative comparison against experimental results. In fact, predicting the binding sites of ssDNA molecules with the metal and the resultant orientation of these molecules is still somewhat controversial; e.g. adenine binding orientation \cite{harroun2018controversial} or orientation dependence due to interaction with magnesium cations \cite{garcia2018} (which are used in our experiments), is expected to lead to considerable variation in the corresponding spectra. 

This work is structured as follows. First, we present DFT simulation results which compare Raman spectra of ssDNA 3-mers under nucleobases permutation with and without non-resonant CE effect. We then present experimental results of the ssDNA 3-mers with an increasing number of enhancement effects; first we present Raman spectra acquired in bulk solution without EM and CE effects, then we present SERS spectra on nanorod array substrates with only EM effect, and finally SERS spectra when both EM and CE effects are present. We analyze the numerical and experimental data, determine prominent trends, and distinguish the spectra of permuted ssDNA 3-mers based on the correlation values of the elements in the corresponding statistical correlation matrix.

\section*{Results}

\subsection*{Simulation results}

\begin{figure}
	\includegraphics[scale=0.25]{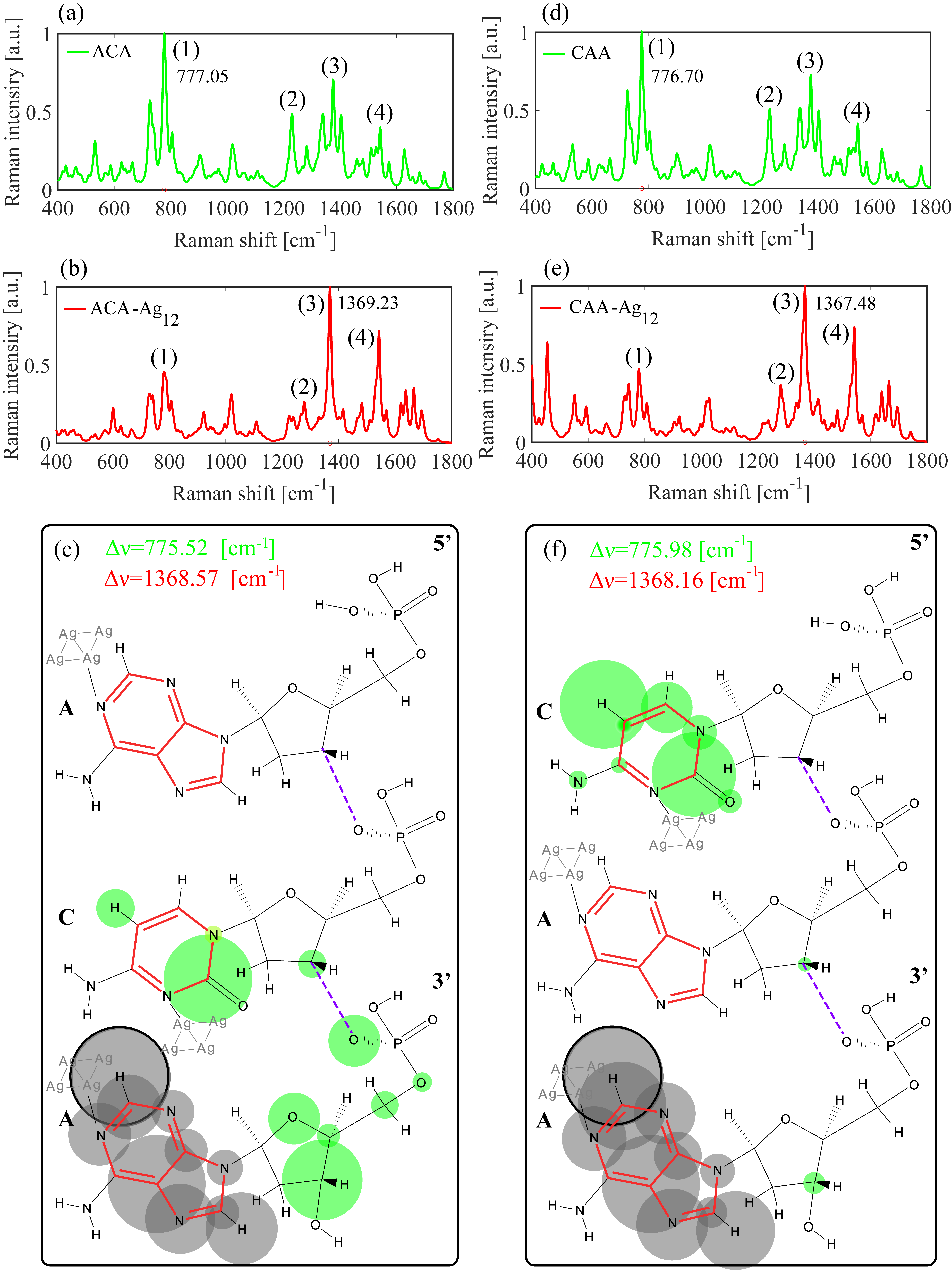}
    \caption{DFT simulation results comparing between Raman spectra of ACA and CAA isomers (group 1) with and without bonded three Ag$_{4}$ nanoparticles. 
    (a,b) Present Raman spectra of ACA and of ACA-Ag$_{12}$ molecules with dominant modes at $775.52$ cm$^{-1}$ and $1368.57$ cm$^{-1}$, respectively, 
whereas (d,e) present Raman spectra of CAA and CAA-Ag$_{12}$ molecules with dominant modes at $775.98$ cm$^{-1}$ and $1368.16$ cm$^{-1}$, respectively.
(c,f) Relative kinetic energy distribution among the different atoms at the dominant normal modes with red disks and green disks corresponding to the cases with and without silver nanoparticles, respectively. The disks with black contours in ACA and CAA molecules admit kinetic energy values larger by a factors $1.22$ and $1.60$, respectively.}
    \label{ACA_CCA}
\end{figure}

\begin{figure}
	\includegraphics[scale=0.25]{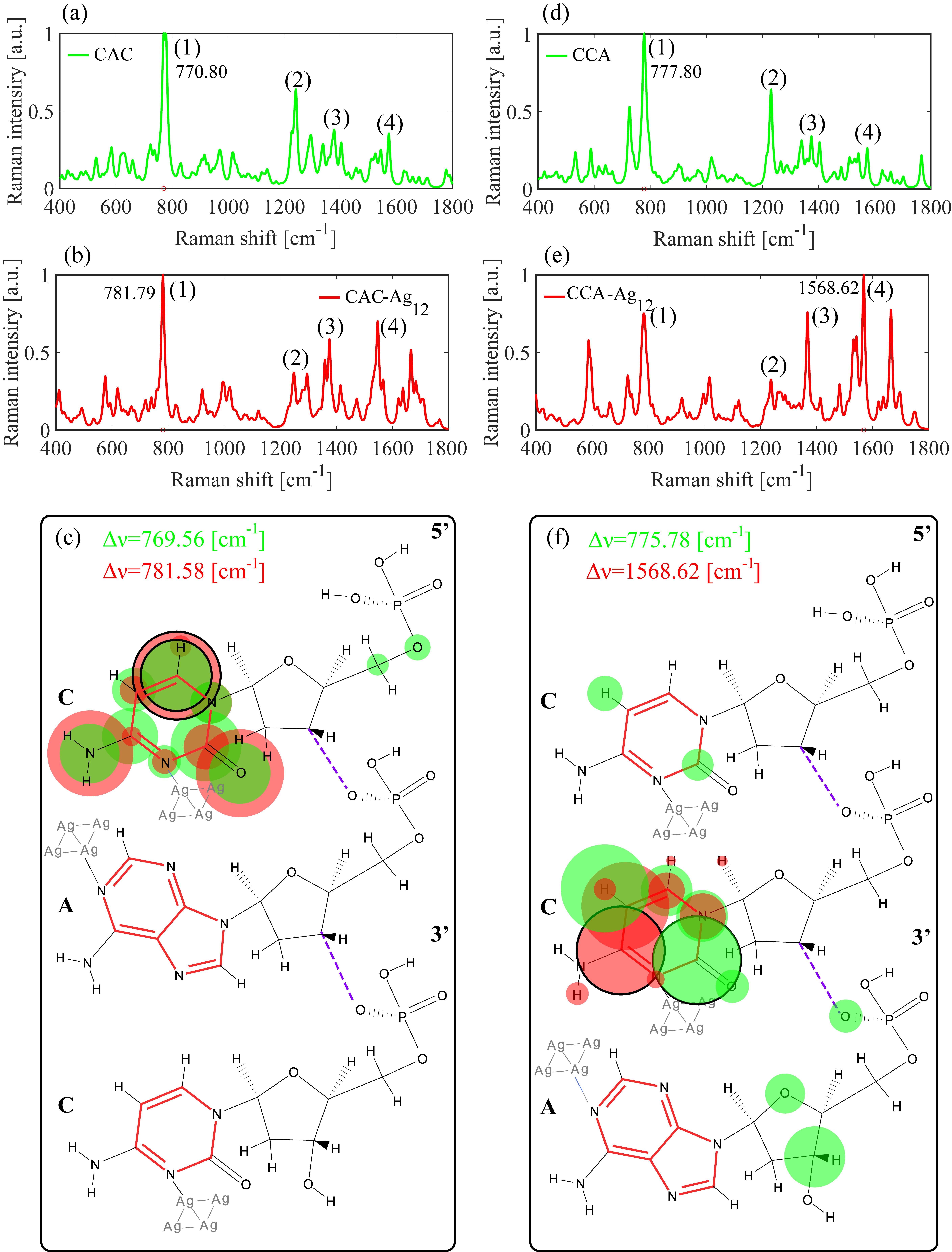}
    \caption{DFT simulation results comparing Raman spectra between CAC and CCA isomers (group 2) with and without three Ag$_{4}$ nanoparticles. 
    (a,b) Present Raman spectra of CAC and of CAC-Ag$_{12}$ at dominant modes $769.56$ cm$^{-1}$ and $781.58$ cm$^{-1}$, respectively, 
whereas (d,e) present Raman spectra of CCA and CCA-Ag$_{12}$ with dominant modes at $775.78$ cm$^{-1}$ and $1568.62$ cm$^{-1}$, respectively.
(c,f) Relative kinetic energy distribution among the different atoms of the dominant modes with red disks and green disks corresponding to the cases with and without silver nanoparticles, respectively. The disks with black contours around red and green disks in CAC molecule indicate atoms with kinetic energy $2.46$ and $1.85$ larger, respectively, whereas in CCA molecules admit kinetic energy values larger by a factors $2.19$ and $1.54$, respectively.}
    \label{CAC_CCA}
\end{figure}

Fig.\ref{ACA_CCA} and Fig.\ref{CAC_CCA} present DFT simulation results of Raman spectra of ACA, CAA (group 1) and CCA, CAC (group 2) sequences, as well as the same sequences where each DNA base is also bound to Ag$_{4}$ nano-particle. All calculated Raman intensity results are normalized to unity, and Raman intensities of the cases with metal particles admit values higher by approximately factor of five or smaller. The broadening of the peaks is manually introduced by Lorentzian functions with bandwidth full-width at half maxima (FWHM) of $12$ cm$^{-1}$, where $5-20$ cm$^{-1}$ is usually employed to describe various broadening effects \cite{le2008principles}.
Fig.\ref{ACA_CCA}(c,f) and Fig.\ref{CAC_CCA}(c,f) present the corresponding 2D schematic representation of the 3-base long sequences where 
each nucleotide consists of a phosphate backbone joined to a sugar (a.k.a. 2-deoxyribose due to a missing hydroxyl) to which the base is attached, 
and different nucleotides are linked through the phosphodiester linkage \cite{WatsonDNA}. The latter also leads to chiral and polar structure of the ssDNA molecule with distinct ends labeled as 3$^{\prime}$ and 5$^{\prime}$. The red/green disks represent the relative kinetic energy of each atom with/without CE effect at the corresponding dominant vibrational mode; solid contour lines around few of the atoms indicate that the relevant atom carries larger kinetic energy as described in figures' captions.
%each nucleotide consists of a phosphate backbone joined to a sugar (a.k.a. 2-deoxyribose due to a missing hydroxyl) 
%at position $2^{\prime}$)
%to which the base is attached \cite{WatsonDNA}; 
%the phosphodiester bond links between the phosphate group of one of the nucleotides closer to the $3^{\prime}$-end to sugar ring of the nucleotide closer to the $5^{\prime}$-end, which defines the chirality (and polarity) of the ssDNA molecule. 
We follow the common convention and write the sequences from $5^{\prime}$-end to $3^{\prime}$-end \cite{WatsonDNA}, where for instance, CAA stands for $5^{\prime}-$CAA$-3^{\prime}$.
%the sequences are represented with $5^{\prime}$-end on top and $3^{\prime}$-end on bottom. 

Our simulation results indicate that the modes can be classified according to the groups that store the kinetic energy of the vibrating atoms, and that the normal frequencies of these groups depend both on their position in the sequence as well as on the closest neighbor DNA bases.  
For instance, the frequencies and the RAs of the ring breathing modes (RBMs) of adenine or cytosine bases, which involve relatively small number of atoms, depend on their location in the sequence and are affected by the DNA bases permutation as we discuss in the following.
Furthermore, changing the number of DNA bases leads to frequency splitting of the normal modes due to partial degeneracy lifting which stems from normal modes dependence on the position along the sequence.

Fig.\ref{ACA_CCA}(a,b,d,e) present Raman spectra of ACA and CAA molecules, group (1), showing intensification, decline and shift of the normal modes, where red and green colors indicate the cases with and without Ag$_{4}$ nanoparticles, respectively. To demonstrate basic trends and patterns we analyze the Raman spectra in the following four spectral regions: 
(1) $775-791$ cm$^{-1}$, (2) $1231-1238$ cm$^{-1}$, (3) $1367-1377$ cm$^{-1}$ and (4) $1541-1544$ cm$^{-1}$, and use the spectra presented in Fig.\ref{ACA_CCA} in conjunction with Table.\ref{TableACA} summarizing modes' frequencies, RAs, as well as the corresponding Raman active group in the sequence. Note that in some cases the dominant peak is composed of several peak of comparable Raman intensity, and therefore the value of the maximal frequency presented in Fig.\ref{ACA_CCA}(a,b,d,e) and Fig.\ref{CAC_CCA}(a,b,d,e) is different than the frequency of the corresponding dominant modes analyzed in Fig.\ref{ACA_CCA}(c,f) and Fig.\ref{CAC_CCA}(c,f).

\textit{Spectral region (1):} is dominant for ACA and CAA molecules due to normal modes which involve 3$^{\prime}$sugar, central (cen) sugar and the single C base. For ACA the dominant frequency at $777.05$ cm$^{-1}$, is mostly due to $775.52$ cm$^{-1}$ (RA$=39.89$ \r{A}$^{4}$/amu) and $779.09$ cm$^{-1}$ (RA$=30.32$ \r{A}$^{4}$/amu) modes. Both admit kinetic energy concentrated at 3$^{\prime}$sugar and C with different out-of plane and ring breathing amplitude vibrations
%symmetric stretching mode 
(see fig.\ref{ACA_CCA}(c) for kinetic energy distribution), whereas kinetic energy of $775.52$ cm$^{-1}$ mode is also concentrated at central sugar group.
Interestingly, as we will see in the following, normal modes with kinetic energy more concentrated mostly at one group are more prone to splitting, i.e. appearance of similar modes with closely separated frequencies at different groups.

In CAA molecule, which is obtained from ACA by  
%From fig.\ref{ACA_CCA}(d) we learn that the conformational change due to 
permutation of $5^{\prime}$A with C, a slight red shift is observed at the peak vibrational Raman intensity to $776.80$ cm$^{-1}$, mostly due to red shift of the $775.02$ cm$^{-1}$ mode to $775.98$ cm$^{-1}$ (RA$=37.36$ \r{A}$^{4}$/amu) and to the decrease of RA of the $779.09$ cm$^{-1}$ mode. 
Bonding to Ag$_{4}$ nanoparticles results in a red shift of all modes in both ACA and CAA molecules, slight decrease of some modes RA (e.g. $779.09$ cm$^{-1}$) and more significant decrease of $778.08$ cm$^{-1}$ RA mode in CAA. 

\textit{Spectral region (2):} describes asymmetric RBM vibrations of the single C base and the attached sugar molecule. Permutation of $5^{\prime}$A with C introduces a relatively minor change in the vibrational frequency from $1231.00$ cm$^{-1}$ to $1231.51$ cm$^{-1}$ and to a slight increase in the corresponding RA (from $47.57$ \r{A}$^{4}$/amu to $50.51$ \r{A}$^{4}$/amu). Bonding to Ag$_{4}$ nanoparticles leads to a blue-shift of the resonant frequency and to a decrease in the RA values.

\textit{Spectral region (3):} describes two normal modes of two different A bases, with practically identical atomic displacements, but with different frequencies due to different position in the sequence. In particular, ACA admits second dominant peak at frequency $1375.35$ cm$^{-1}$ which is a contribution of the following stretching-bending modes (which lead to change in bonds' length and angles): $3^{\prime}$A at $1374.37$ cm$^{-1}$ (RA$=59.78$ \r{A}$^{4}$/amu) and similar mode of $5^{\prime}$A at $1376.74$ cm$^{-1}$ (RA$=50.11$ \r{A}$^{4}$/amu). From Table.\ref{TableACA} below we learn that under permutation these modes are concentrated at A bases, but with slightly different values of frequency and RA. Under bonding to Ag$_{4}$ nanoparticles all modes are intensified, slightly red shifted, and the vibrational modes become concentrated also in the adjacent sugar groups.

\textit{Spectral region (4):} presents vibrational modes of two A bases around $1542$ cm$^{-1}$ with two slightly different frequencies.  
%with low RA, which serves as a ``seeded mode'' for vibrational mode with high RA which is switched on once Ag$_{4}$ particles are bond to the molecule as discussed below. 
Permutation of $5^{\prime}$A with C in ACA molecule leads to slight intensification of RA and  red shift of the vibrational frequencies, whereas bonding to Ag$_{4}$ nanoparticles leads to more prominent intensification of the RA values and red shift of practically all modes. 

Due to the chiral nature of DNA molecule, AAC sequence, obtained by permuting C base with 3$^{\prime}$A base in CAA sequence, is expected to admit a different Raman spectra.
Indeed in Supplementary Information we describe additional DFT simulation results indicating a red shift of cytosine active mode in the second spectral region compared to ACA and CAA sequences, and also lower RA values of adenine modes in the fourth spectral region when bound to Ag$_{4}$ nanoparticles.

%both intensification and diminishmet of RA values of some modes, accompanied by spectral shift of the corresponding resonant frequencies. Specifically, Fig.\ref{ACA_CCA}(b) indicates that Ag$_{4}$ particles bond to nitrigenous rings lead to intesification and slight blue shift of the asymmetric RBMs of A bases in the spectral region (2); frequency shift from $1374.37$ cm$^{-1}$ and  $1376.74$ cm$^{-1}$, to $1368.57$ at $3^{\prime}$ A and to $1371.40$ at $5^{\prime}$ (see Table.\ref{TableACA} for RA values). Furthermore, Ag$_{4}$ bonding leads to slight increase of the asymetric RBM mode of C at $779.09$ cm$^{-1}$ (RA$=30.3239$ \r{A}$^{4}$/amu) mentioned above to a red-shifted frequency $790.48$ cm$^{-1}$. 
%Finally, Fig.\ref{ACA_CCA}(e) indicates that $5^{\prime}$ A and C bases permutation leads to slight blue shift of the dominant breathing modes of A to $1367.45$ cm$^{-1}$ $1368.16$ cm$^{-1}$ and significant increase of RA from RA$=90.96$ \r{A}$^{4}$/amu to RA$=201.37$ \r{A}$^{4}$/amu.   

\begin{table}
\begin{tabular}{ccccc}

\\[-0.5 ex] 
\textcolor{white}{Name}&  
\begin{tabular}{@{}c@{}}  Frequency [cm$^{-1}$]   \end{tabular} & 
\begin{tabular}{@{}c@{}}  Raman Activity [\r{A}$^{4}$/amu]  \end{tabular} & 
\begin{tabular}{@{}c@{}}  Mode location   \end{tabular} 
\\[1 ex] 

\hline
ACA&  
\begin{tabular}{@{}c@{}} $767.08$ \quad \\ $773.34; 775.52; 779.09$ \quad \\ $1231.00$ \quad \\ $1374.37; 1376.74$ \quad \\ $1541.24; 1542.54$ \end{tabular} & 
\begin{tabular}{@{}c@{}} $2.49$ \quad \\ $6.91; 39.88; 30.23$ \quad \\ $47.57$ \quad \\ $59.78; 50.11$ \quad \\ $28.11; 44.66$ \end{tabular} & 
\begin{tabular}{@{}c@{}} C \quad \\ C; 3$^{\prime}$ sugar, cen sugar, C; C, $3^{\prime}$ sugar \quad \\ C, cen sugar \quad \\ $3^{\prime}$A; $5^{\prime}$A \quad \\ $5^{\prime}$A; $3^{\prime}$A \end{tabular} 
\\ [1ex]

\hline
CAA&  
\begin{tabular}{@{}c@{}} $766.82; 778.08$ \quad \\ $775.98; 776.94$ \quad \\ $1231.51$ \quad \\  $1374.52; 1376.57$ \quad \\ $1541.86; 1542.66$ \end{tabular} & 
\begin{tabular}{@{}c@{}} $6.47; 18.45$ \quad \\ $37.36; 17.45$ \quad \\ $50.51$ \quad \\ $57.40; 54.16$ \quad \\ $29.33; 47.09$ \end{tabular} & 
\begin{tabular}{@{}c@{}}  C; C \quad \\ C, 3$^{\prime}$sugar, 3$^{\prime}$PHB; C, 3$^{\prime}$sugar, 3$^{\prime}$PHB \quad \\ C, cen sugar \quad \\ 3$^{\prime}$A; cen A \quad \\ cen A; $3^{\prime}$A \end{tabular} 
\\ [1ex]

\hline
ACA-Ag$_{12}$&  
\begin{tabular}{@{}c@{}}  $783.65; 790.48$ \quad \\ $1237.93$ \quad \\ $1368.57; 1371.40$ \quad \\ $1542.92; 1543.55$  \end{tabular} & 
\begin{tabular}{@{}c@{}}  $23.76; 38.14$ \quad \\ $21.12$ \quad \\ $208.06; 110.78$ \quad \\ $149.83; 106.66$   \end{tabular} & 
\begin{tabular}{@{}c@{}}  C; C \quad \\ C, cen sugar \quad \\ $3^{\prime}$A, $3^{\prime}$sugar; $5^{\prime}$A, $5^{\prime}$sugar \quad \\ $3^{\prime}$A;  $5^{\prime}$A  \end{tabular} 
\\ [1ex]

\hline
%\\
%\\[-6.5ex] 
CAA-Ag$_{12}$& 
\begin{tabular}{@{}c@{}} $778.75; 781.09$ \quad \\ $1237.37$ \quad \\ $1367.45; 1368.16$ \quad \\ $1541.41; 1543.41$ \end{tabular} & 
\begin{tabular}{@{}c@{}} $58.46; 4.48$ \quad \\ $25.69$ \quad \\ $90.96; 201.37$ \quad \\ $152.31; 147.26$ \end{tabular} & 
\begin{tabular}{@{}c@{}} 3$^{\prime}$sugar; C  \quad \\ C, cen sugar \quad \\ cen A, cen sugar; 3$^{\prime}$A, 3$^{\prime}$sugar \quad \\ cen A; 3$^{\prime}$A \end{tabular}
\\ 
\\ [-2.5ex]
\hline
\end{tabular}    
\caption{DFT simulation results presenting the effect of C-3$^{\prime}$A permutation as well as bases bonding to Ag$_{4}$ particles on the frequency and the Raman activity of the normal modes in the four different spectral regions: $775-791$ cm$^{-1}$, $1231-1238$ cm$^{-1}$, $1367-1377$ cm$^{-1}$ and $1541-1544$ cm$^{-1}$.}
\label{TableACA}
\end{table}

Consider now group (2) with sequences CAC and CCA, which are formed by replacing one of the A bases in group (1) by a C basis. Following our discussion above, we expect normal modes splitting for the two C bases and less modes involving a single A base. Fig.\ref{CAC_CCA}(a,b,d,e) present Raman spectra of CAC and CCA molecules, and similarly to the discussion above for group (1), we consider the following four spectral regions: 
(1) $766-790$ cm$^{-1}$, (2) $1230-1242$ cm$^{-1}$, (3) $1367-1375$ cm$^{-1}$ and (4) $1542-1574$ cm$^{-1}$. We use Fig.\ref{CAC_CCA} results in conjunction with Table.\ref{TableCAC} summarizing modes' frequencies, RAs, as well as the active group in the sequences.

\textit{Spectral region (1):} describes mode splitting due to a pair of C bases which introduces additional normal modes. In particular, the modes $767.08$ cm$^{-1}$ in ACA sequence and $766.82$ cm$^{-1}$ in CAA sequence split into the pair $766.33, 769.56$ cm$^{-1}$ in CAC sequence and into the pair $764.82, 766.11$ cm$^{-1}$ in the CCA sequence (see Table.\ref{TableCAC}). Furthermore, the pair $775.98$ cm$^{-1}$ and $776.94$ cm$^{-1}$, which involves several groups in CAA sequence, joins into another mode which involves several molecular groups at frequency $779.33$ cm$^{-1}$. Bonding to Ag$_{4}$ nanoparticles leads to a drop in RA, especially for the $769.56$ cm$^{-1}$ mode in CAC sequence.

\textit{Spectral region (2):} describes mode splitting of the $1231.00$ cm$^{-1}$ and $1231.51$ cm$^{-1}$ C-based modes in ACA and CAA sequences (with single C), respectively, to the C-based paired modes $1241.62$ cm$^{-1}$, $1242.74$ cm$^{-1}$ and $1230.04$ cm$^{-1}$, $1231.15$ cm$^{-1}$ in CAC and CCA sequences (with two C's), respectively. Bonding to Ag$_{4}$ nanopartices leads to decrease of all RAs except for the $1242.74$ cm$^{-1}$ mode in CAC sequence which increases from $17.47$ \r{A}$^{4}$/amu to $39.20$ \r{A}$^{4}$/amu.

\textit{Spectral region (3):} describes the activity of a single Raman active A-based normal mode, as opposed to a pair of similar A-based modes in group (1). Under permutation from CAC to CCA the resonant frequency is red-shifted from $1378.61$ cm$^{-1}$ to $1374.32$ cm$^{-1}$ with slight increase in RA as further described in Table.\ref{TableCAC}. Bonding to Ag$_{4}$ nanoparticles leads to red shift and intensification of this mode in both CAC and CCA sequences (see Table.\ref{TableCAC}).

\textit{Spectral region (4):} Under permutation of 3$^{\prime}$C with A, RA of A mode in CAC sequence increases from $35.28$ \r{A}$^{4}$/amu to $44.27$ \r{A}$^{4}$/amu in CCA sequence, and to decrease of RA values of the C-based pair modes. Bonding to Ag$_{4}$ nanoparticles leads to strong intensification of both modes in the CCA-Ag$_{12}$ sequence. In particular, RA of cen C mode in CCA sequence with RA$=29.74$ \r{A}$^{4}$/amu increase to RA$=282.20$ \r{A}$^{4}$/amu in CCA-Ag$_{12}$ sequence. 

\begin{table}
\begin{tabular}{ccccc}

\\[-0.5 ex] 
\textcolor{white}{Name}&  
\begin{tabular}{@{}c@{}}  Frequency [cm$^{-1}$]   \end{tabular} & 
\begin{tabular}{@{}c@{}}  Raman Activity [\r{A}$^{4}$/amu]  \end{tabular} & 
\begin{tabular}{@{}c@{}}  Mode location   \end{tabular} 
\\[1 ex] 

\hline
CAC&  
\begin{tabular}{@{}c@{}}  $766.33; 769.56$ \quad \\ $773.42; 773.78$ \quad \\ $779.33$ \quad \\ $1241.62; 1242.74$ \quad \\ $1378.61$  \quad \\ $1545.30; 1572.78; 1573.49$
\end{tabular} & 
\begin{tabular}{@{}c@{}}  $7.25; 45.43$ \quad \\ $7.34; 1.45$ \quad \\ $53.45$  \quad \\ $53.46; 17.47$ \quad \\ $45.21$ \quad \\ $35.28; 36.30; 32.40$ 
\end{tabular} & 
\begin{tabular}{@{}c@{}}  5$^{\prime}$C; 3$^{\prime}$C \quad \\ 3$^{\prime}$C; 5$^{\prime}$C  \quad \\ 3$^{\prime}$sugar, cen sugar, cen PHB, 3$^{\prime}$PHB, 3$^{\prime}$sugar \quad \\ 5$^{\prime}$C, 5$^{\prime}$sugar; 3$^{\prime}$C, 3$^{\prime}$sugar   \quad \\ A \quad \\ A; $3^{\prime}$C; $5^{\prime}$C \end{tabular} 
\\ [1ex]

\hline
CCA&  
\begin{tabular}{@{}c@{}}  $764.82; 766.11$ \quad \\ $775.78$ \quad \\ $778.20; 779.45$ \quad \\ $1230.04; 1231.15$ \quad \\ $1374.32$ \quad \\ $1542.30; 1572.84; 1573.99$ \end{tabular} & 
\begin{tabular}{@{}c@{}}  $4.62; 5.64$ \quad \\ $36.28$ \quad \\ $21.87; 32.23$ \quad \\ $47.34; 49.62$ \quad \\ $59.14$ \quad \\ $44.77; 33.69; 29.74$ \end{tabular} & 
\begin{tabular}{@{}c@{}}  cen C; 5$^{\prime}$C \quad \\ cen C, 3$^{\prime}$sugar, 3$^{\prime}$PHB, 5$^{\prime}$C \quad \\ 5$^{\prime}$C; cen C, 3$^{\prime}$sugar, 3$^{\prime}$PHB \quad \\ cen C; 5$^{\prime}$C \quad \\ A \quad \\ A; 5$^{\prime}$C; cen C \end{tabular} 
\\ [1ex]

\hline
CAC-Ag$_{12}$&  
\begin{tabular}{@{}c@{}}  $767.73; 769.56$ \quad \\ $781.58; 782.66$ \quad \\ $1245.54; 1250.79$ \quad \\ $1375.63$ \quad \\ $1546.52; 1565.53; 1568.45$ \end{tabular} & 
\begin{tabular}{@{}c@{}}  $2.62; 5.17$ \quad \\ $57.42; 34.49$ \quad \\ $21.48; 39.20$ \quad \\ $116.53$ \quad \\  $119.22; 28.48; 46.86$  \end{tabular} & 
\begin{tabular}{@{}c@{}}  3$^{\prime}$C; 5$^{\prime}$C \quad \\ 5$^{\prime}$C; 3$^{\prime}$C, cen sugar \quad \\ 3$^{\prime}$C, 3$^{\prime}$sugar; 5$^{\prime}$C, 5$^{\prime}$sugar \quad \\ A, cen sugar  \\ A; $3^{\prime}$C; $5^{\prime}$C  \end{tabular} 
\\ [1ex]

\hline
%\\
%\\[-6.5ex] 
CCA-Ag$_{12}$& 
\begin{tabular}{@{}c@{}} $767.95; 769.68$ \quad \\ $783.26; 789.02$ \quad \\ $1237.27; 1238.13$ \quad \\ $1368.33$ \quad \\ $1543.09; 1568.30; 1569.83$ \end{tabular} & 
\begin{tabular}{@{}c@{}} $1.61; 2.38$ \quad \\ $40.66; 45.30$ \quad \\ $30.89; 30.23$  \quad \\ $208.60$ \quad \\ $156.58; 282.20; 69.38$ \end{tabular} & 
\begin{tabular}{@{}c@{}} cen C; 5$^{\prime}$C \quad \\ cen C; cen C, PHB, 5$^{\prime}$C  \quad \\ cen C; 5$^{\prime}$C \quad \\  A, 3$^{\prime}$sugar \quad \\ A; cen C; 5$^{\prime}$C \end{tabular}
\\ 
\\ [-2.5ex]
\hline
\end{tabular}    
\caption{DFT simulation results presenting the effect of C-3$^{\prime}$A permutation and Ag$_{4}$ nanoparticles bonding to the molecules, on the normal modes' frequencies and Raman activities in the four spectral regions: $775-790$ cm$^{-1}$, $1230-1250$ cm$^{-1}$, $1367-1379$ cm$^{-1}$ and $1542-1574$ cm$^{-1}$.}
\label{TableCAC}
\end{table}

Due to chiral nature of DNA molecule ACC sequence, obtained by permuting A base with 5$^{\prime}$C base in CAC sequence, is expected to admit a different Raman spectra.
In Supplementary Information we describe additional DFT simulation results which allow to compare ACC to other sequences. These include blue shift of the cytosine modes in fourth spectral region, $1572.78$ cm$^{-1}$ and $1573.49$ cm$^{-1}$ in CAC sequence to $1573.07$ cm$^{-1}$ and $1576.45$ cm$^{-1}$ in ACC sequence. Furthermore, from Table.\ref{TableCAC} and Table.S1 we learn that cytosine modes, $1568.30$ cm$^{-1}$ and $1569.83$ cm$^{-1}$, in CCA-Ag$_{12}$ sequences experience blue shift to $1570.39$ cm$^{-1}$ and $1572.39$ cm$^{-1}$ in ACC-Ag$_{12}$ sequences accompanied with a decrease in RA. 

\textcolor{black}{Importantly, as indicated in Fig.S1, each one of the ssDNA molecules ACA, CAA, CAC, and CCA admit different spatial conformation. The different conformations within each one of the groups (1) and (2), due to permutation modified intramolecular interaction, leading to different distances between the atoms and eventually to modification of normal modes and Raman/SERS spectra. We then assume that the optimized geometry holds also for the case when the sequences are bond to metal nanolusters, somewhat mimicking behavior of real molecules acquiring a given shape prior to metal adsorption. Metal atoms further modify polarizability properties of the ssDNA molecules leading to modification of the SERS/Raman signal and the kinetic energy distribution of the dominant modes as indicated above.}

\subsection*{Experimental results}

\begin{figure}[h]
	\includegraphics[scale=0.53]{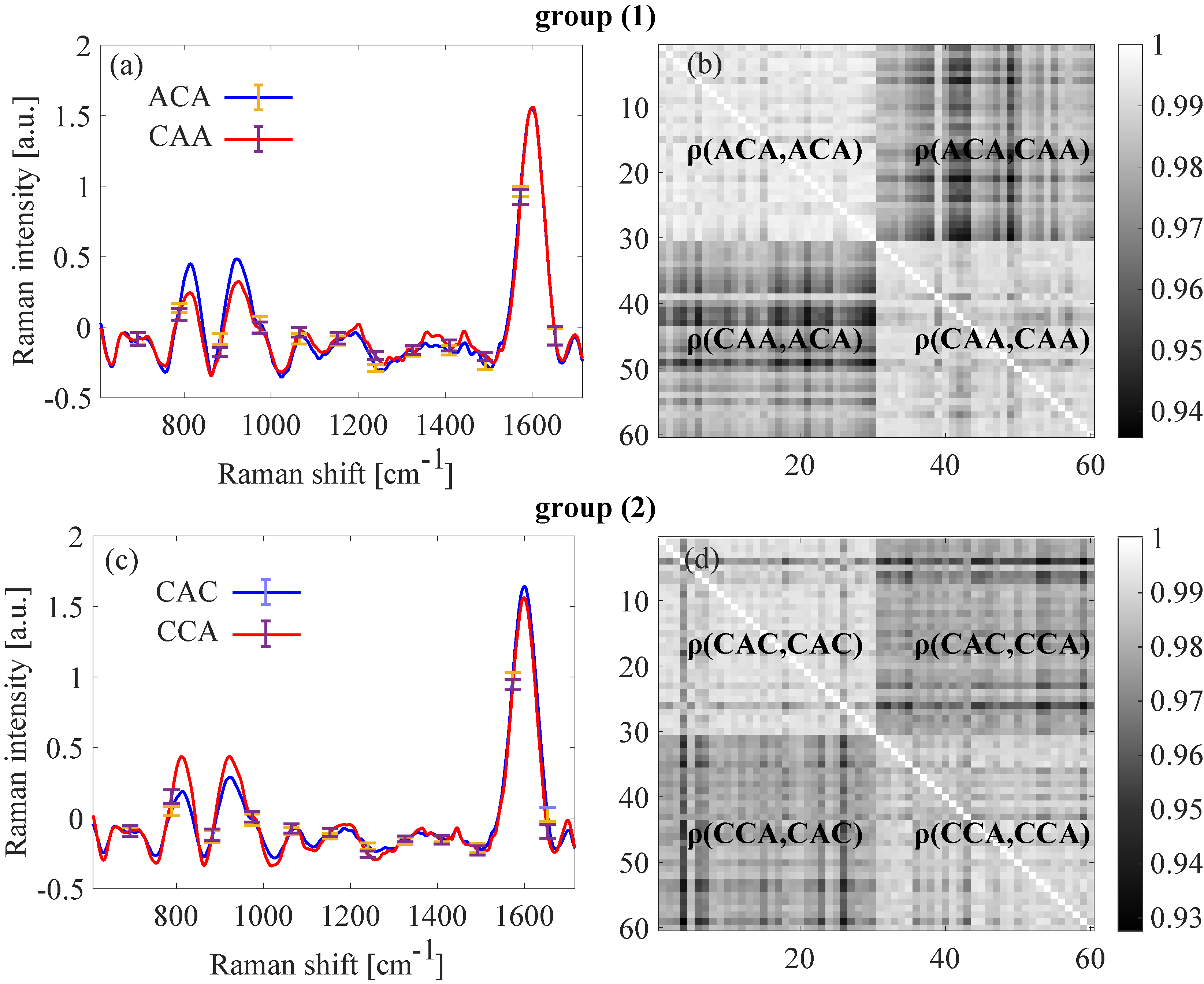}
    \caption{Experimental results presenting mean Raman spectra and correlation matrices of $30$ measurements of ACA, CAA, CAC and CCA, ssDNA molecules in bulk solution. (a) Mean Raman spectra of ACA (blue) and CAA (red), and (b) presents the corresponding statistical correlation matrix. 
%     indicating mostly appearance of vibrational peak around $1036$ cm$^{-1}$ in ACA spectra.  
  (c) Mean Raman spectra of CAC (blue) and CCA (red), and (d) presents the corresponding statistical correlation matrix.}
    \label{CovarianceRamanNoMetal}
\end{figure}

\begin{figure}
	\includegraphics[scale=0.53]{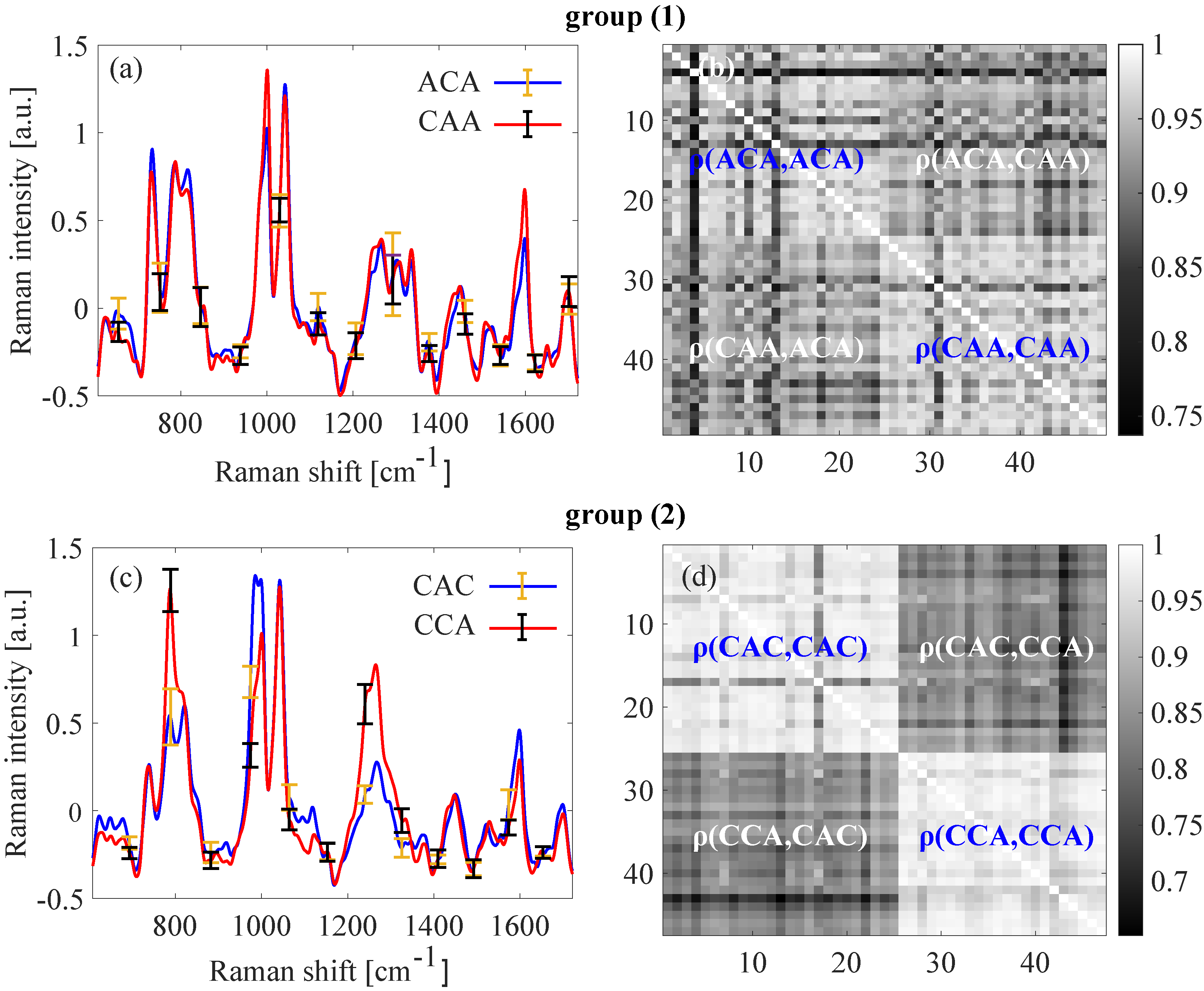}
    \caption{Experimental results presenting the effect of permutation of DNA bases, 
    adsorbed to nanorod gold substrate covered with $2$ nm thick Al$_{2}$O$_{3}$ film, on the mean SERS spectra and correlation matrices without the CE effect. The experimental set is comprised of $24$, $25$, $25$ and $22$ measurements of ACA, CAA, CAC and CCA, ssDNA molecules, respectively. (a) Mean SERS spectra of ACA (blue) and CAA (red), and (b) presents the corresponding correlation matrix. (c) Mean SERS spectra of CAC (blue) and CCA (red) and (d) presents the corresponding correlation matrix.}
    \label{CovarianceALD}
\end{figure}

\begin{figure}
	\includegraphics[scale=0.53]{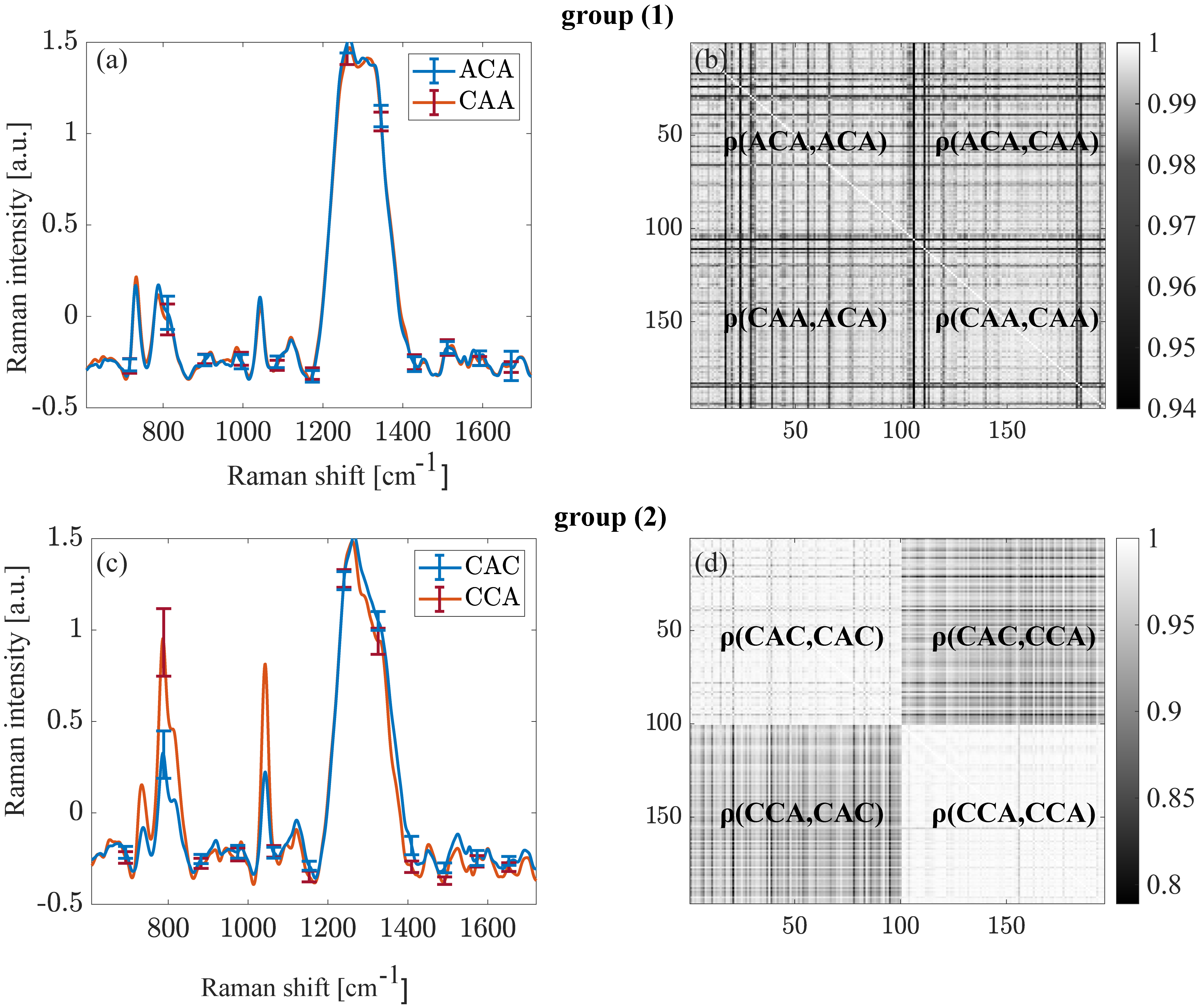}
    \caption{Experimental results presenting the effect of permutation of DNA bases,
    adsorbed to nanorod silver substrate covered with $2$ nm thick Al$_{2}$O$_{3}$ film, on the mean SERS spectra and correlation matrices without the CE effect. The experimental set is comprised of $98$, $98$, $100$ and $96$ measurements of ACA, CAA, CAC and CCA, ssDNA molecules, respectively. (a) Mean SERS spectra of ACA (blue) and CAA (red), and (b) presents the corresponding correlation matrix. (c) Mean SERS spectra of CAC (blue) and CCA (red) and (d) presents the corresponding correlation matrix.}
    \label{CovarianceALDSilver}
\end{figure}

\begin{figure}
	\includegraphics[scale=0.53]{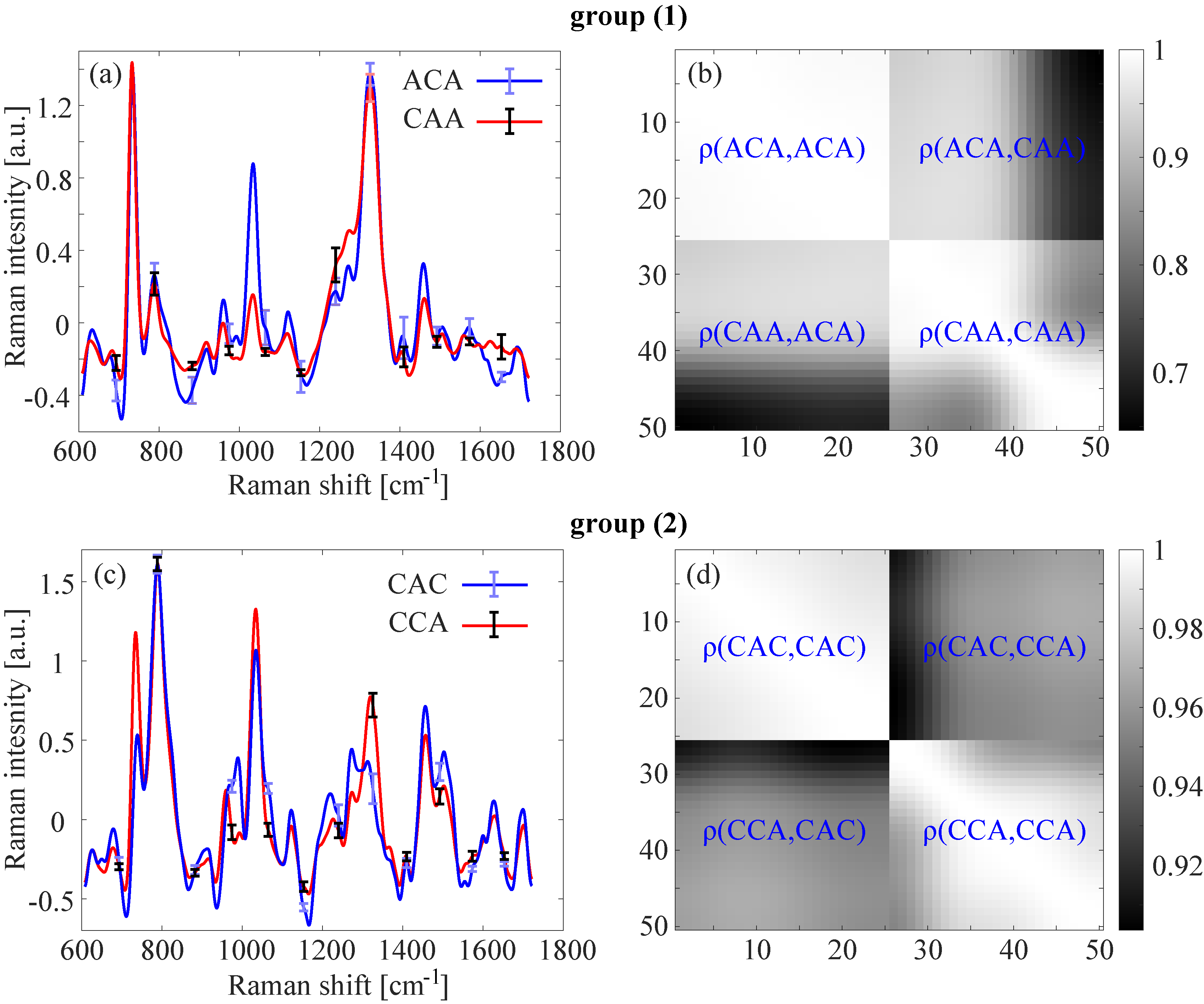}
    \caption{Experimental simulation results presenting mean SERS spectra of $25$ measurements of each ssDNA molecule adsorbed to nanorod silver substrate and the corresponding statistical correlation matrices between the different measurements. (a) Mean SERS spectra of ACA (blue) and CAA (red), and (b) presents the corresponding statistical correlation matrix. 
%     indicating mostly appearance of vibrational peak around $1036$ cm$^{-1}$ in ACA spectra.  
   Minimal values in the first and second diagonal block of the statistical correlation matrix are $0.991$ and $0.817$, whereas the off-diagonal blocks admit lowest values $0.646$. 
   (c) Mean SERS spectra of CAC (blue) and CCA (red). (d) Minimal values in the first and second diagonal block of the diagonal matrix are $0.987$ and $0.954$, whereas the off-diagonal blocks admit lowest values $0.903$.}
    \label{Covariance}
\end{figure}

Fig.\ref{CovarianceRamanNoMetal} and Figs.\ref{CovarianceALD}-\ref{Covariance}, respectively, present experimental Raman and SERS measurement results under exciting laser of wavelength $785$ nm, where each measurement is conducted separately on each one of the following molecules ACA, CAA, CAC, CCA; the first and the second pairs are referred below as group (1) and group (2), respectively.
First, we consider Raman signal of ssDNA molecules in IDTE bulk solution, i.e. without metal substrate and therefore without the accompanying EM and CE effects. 
\textcolor{black}{The order of magnitude of the EM enhancement factor, as defined according to \citep[eq.4]{leru2007}, which compares the total SERS signal from a given surface element per estimated number of molecules adsorbed to the illuminated surface area, to the total Raman signal from a given spatial illuminated volume (defined by beam waist and Rayleigh length) per number of molecules in that volume, is $\sim 10^{2}$}. Fig.\ref{CovarianceRamanNoMetal}(a,b) and Fig.\ref{CovarianceRamanNoMetal}(c,d) present comparison between Raman spectra of group (1) and between group (2) sequences, respectively, where for clarity of presentation range bars are presented at 12 points along the profile and indicate 95\% confidence intervals on the mean.
Importantly, here and elsewhere unless otherwise specified, the experimental data was subject to multiplicative-scattering  correction (MSC) in order to filter out various noises, such as multiplicative light  scattering \cite{Isaaksson1988}. 
In fact, unless ssDNA molecules directly adsorbed to the metal, MSC is necessary to observe the emerging square regions in the correlation matrices Fig.\ref{CovarianceRamanNoMetal}(b,d) (as well as in Fig.\ref{CovarianceALD}(b,d) and in Fig.\ref{CovarianceALDSilver}(b,d)). The latter are defined as the ratio of the covariance of the two variates to the variance product of the independent variates \cite{GibbonsChakraborti}, admit
close to unity values in the self-correlation blocks (e.g. $\rho(\text{ACA},\text{ACA})$, $\rho(\text{CAA},\text{CAA})$, and lower values in the cross-correlation blocks (e.g. $\rho(\text{ACA},\text{CAA})$), signaling a difference in Raman spectra due to nucleobases permutation. 

Next we consider the case where ssDNA molecules are adsorbed to $2$ nm thick Al$_{2}$O$_{3}$ dielectric film deposited on top of oblique-angled deposition (OAD) \cite{barranco2016} fabricated gold nanorod arrays. In so doing we introduce EM effect due to gold nanorod array but eliminate the resonant charge transfer CE effect.
Figs.\ref{CovarianceALD}(a,c) present mean SERS spectra of ACA and CAA sequences, respectively, accompanied by the corresponding statistical correlation matrices Fig.\ref{CovarianceALD}(b,d). 
The latter indicate a close to unity correlation in the diagonal blocks and smaller correlation in the off-diagonal block which is more visible between group (2) sequences. 
In order to eliminate charge transfer CE effect, we perform similar experiments with ssDNA adsorbed to $2$ nm thick Al$_{2}$O$_{3}$ dielectric film deposited on top of silver OAD nanorod arrays. Fig.\ref{CovarianceALDSilver} presents the corresponding experimental results where Figs.\ref{CovarianceALDSilver}(a,c) illustrate the mean spectra of group (1) and (2) ssDNA, and Figs.\ref{CovarianceALDSilver}(b,d) present the correlation matrices indicating the differences between the two groups.
Similarly to the case of gold nanorod arrays coated with Al$_{2}$O$_{3}$, presented in Fig.\ref{CovarianceALD} above the observed differences between group (1) molecules are smaller than between group (2) molecules; the results in Fig.\ref{CovarianceALDSilver}(a,b) lead to very close signal and the sequences ACA, CAA adsorbed to silver nanorod array coated with Al$_{2}$O$_{3}$ are practically indistinguishable. Interestingly, the pattern of smaller differences between group (1) molecules than between group (2) molecules is also consistent with the computational spectra which can be seen by visually comparing the mean spectra of group (1) molecules, Fig.\ref{ACA_CCA}(a) vs Fig.\ref{ACA_CCA}(d), and Fig.\ref{ACA_CCA}(b) vs Fig.\ref{ACA_CCA}(e); group (2) molecules, Fig.\ref{CAC_CCA}(a) vs Fig.\ref{CAC_CCA}(d), and Fig.\ref{CAC_CCA}(b) vs Fig.\ref{CAC_CCA}(e) due to more pronounced changes of the $5^{\prime}$C relative to $3^{\prime}$A due to A-C permutation of the nearest nucleobase.

Next we consider the case where ssDNA molecules are adsorbed to silver nanorod arrays which support both CE and EM effects.
Figs.\ref{Covariance}(a,c) present SERS spectra of the four ssDNA sequences with similar range bars as presented in figures above. Fig.\ref{Covariance}(b,d) present the corresponding statistical correlation matrices indicating close to unity correlation in the diagonal blocks and smaller correlation in the off-diagonal blocks. In fact, in this case MSC wasn't implemented and signal processing wasn't necessary to observe the difference between the permuted cases. Interestingly, unlike the case presented in Fig.\ref{CovarianceALD} and Fig.\ref{CovarianceALDSilver} above, where differences between group (1) sequences are less visible compared to differences between group (2) sequences, in this case the differences between group (2) sequences is more visible compared to group (1) sequences. We attribute this difference to more prominent CE effect of adenine bases reported in the recent work \cite{Nguyen2020}. 

%1030 cm-1 ag4 simulation

\subsection*{Methods}

\subsubsection*{DFT simulation setup and governing equations}

In this work we employ time-dependent functional density theory (DFT) by using Gaussian $16$ software \cite{gaussian} at San Diego Supercomputer Center (SDSC) at the University of California, San Diego \cite{XSEDE}) by implementing a built-in Becke’s three-parameters hybrid functionals and Lee-Yang-Parr (B3LYP) electron correlation; $6-31G^{*}$ basis set is used for H, C, P, O, N atoms and LANL2DZ basis set is used for Ag atoms to achieve better accuracy. 
All simulated ssDNA molecules were electrically neutral singlets, where additional protons bonded to the oxygen atoms in the phosphate backbone screened the negative charge of the phosphate group.

The Raman cross section of mode $k$, $\left( d\sigma/d\Omega \right)_{k}$, and the corresponding Raman intensity, $I$, are given by 
\begin{subequations}
\begin{align}
	\left( \dfrac{d \sigma}{d \Omega} \right)_{k} &= f \dfrac{(\nu_{L}-\nu_{k})^{4}}{\nu_{k}} B_{k}(T) S_{k},
\\	
	I &= \sum_{k} \left( \dfrac{d \sigma}{ d \Omega} \right)_{k} L(\nu-\nu_{k}).
\end{align}
\end{subequations}
Here, $B_{k}(T)$ and $L(\nu-\nu_{k})$ are the temperature dependent correction and the Lorentzian function centered at $\nu_{k}$ both given by
 \begin{subequations}
 \begin{align}
 	 B_{k}(T) &= 1/\left(1 - e^{-\nu_{k}/\nu_{T}} \right), \quad \nu_{T} \equiv hc/(k_{B}T)
 \\	 
 	 L(\nu-\nu_{k}) &= \dfrac{1}{\pi g} \dfrac{g^{2}}{g^{2}+(\nu-\nu_{k})^{2}},	 
 \end{align}
 \end{subequations}
respectively. Here, $h$, $c$, $k_{B}$, $T$ are the corresponding Planck's constant, speed of light, Boltzmann constant, room temperature, respectively, whereas $g$ is the half-width at half height (HWHH) parameter of the Lorentzian which represents broadening effects. Typically, for room temperature values ($298.13$ K and $\nu_{T}=207.27$ cm$^{-1}$) the temperature dependent factor introduces correction below $10 \%$ for frequencies above $497.01$ cm$^{-1}$.

\subsubsection*{Substrate and sample preparation}

Silver and gold nanorod substrate arrays were fabricated using a single-step OAD method \cite{barranco2016} 
\textcolor{black}{by employing Denton Discovery Sputter system. Nanorods formation is governed by the so-called ‘‘shadowing effect”; when a material flux of deposited particles is made incident on the target surface, initially formed nuclei structures prevent the deposition of later arriving particles in regions situated behind them
(see \cite{barranco2016} and references within), thus leading to growth of only initial nuclei. 
In our case the substrate is oriented at zenithal deposition angle $75^o$ leading to a nanorod tilt angle of $60^o$ relative to the substrate normal. Fig.S2 in the Supplementary Material section describes SEM images of a typical nanorod array used in our experiments.}

The metal nanorod arrays, which were used to eliminate charge trasfer CE effect, were covered with $2$ nm Al$_{2}$O$_{3}$ thick film using Atomic Layer Deposition (ALD); see \cite{Nguyen2020} for fabrication details and also sample preparation steps. Briefly, the ssDNA solutions were prepared by diluting the DNA stock solution in 4-(2-hydroxyethyl)-1- piperazine ethanesulfonic acid (HEPES) and MgCl$_{2}$. We then drop-casted this ssDNA solution onto the metal nanorod substrate, let it dry overnight, rinsed with deionized water in order to remove excess of crystallized salt and unbound ssDNA molecules, and then blow-dried.

For acquisition of Raman spectra of ssDNA molecules suspended in bulk solution we employed high throughput $5$ mm diameter Wilmad NMR tubes filled with ssDNA sequences in IDTE buffer (pH $8.0$, $10$ mM Tris-HCl/$0.1$ mM EDTA) normalized to a concentration of $100$ $\mu$M.

\subsubsection*{SERS and Raman measurement and data processing}

SERS and Raman measurements were acquired with Renishaw inVia Raman spectrometer by employing 
%$50 mW$
\textcolor{black}{$10$ mW} continuous 
%wavelength 
$785$ nm excitation wavelength; acquisition time of $5$ s per spectrum. The objective magnification used was $50X$ with $NA = 0.75$. 
The grating type used was $1200$ l/mm at $785$ nm, and the grating setting was set to a static regime with spectrum range extending between $600$
cm$^{-1}$ to $1700$ cm$^{-1}$ with spectral resolution around $1$ cm$^{-1}$. The mapping setting of the spectrometer was used to acquire $100$ measurements from a total substrate area of dimensions $50 \times 50$ $\mu m^{2}$; this area is divided into $10 \times 10$ square units, where the dimension of
each area unit is $5 \times 5$ $\mu m^{2}$ and each acquired spectrum measurement is taken from a different square area.
\textcolor{black}{Raman measurements of bulk solution (i.e. without metal substrate) were performed by employing $50$ mW continuous $532$ nm excitation wavelength; objective magnification $5X$ with $NA = 0.12$; grating $1800$ l/mm.}

\textcolor{black}{After data acquisition, SERS/Raman spectra undergo the following signal processing steps which include: baseline correction, cosmic ray removal and MSC. The first two were implemented by employing built-in algorithms in Renishaw WiRE 4 software with Gaussian smoothing window of length 20, whereas MSC was implemented in MATLAB codes. These procedures lead to small artifact regions with negative values of Raman intensity, which do not affect the derived statistical correlation matrices.}

\textcolor{black}{The statistical analysis was performed by employing built-in 'corrcoef' function in Matlab (version 9.7.0.1261785 (R2019b), Natick, Massachusetts, The MathWorks Inc.) which calculates Pearson's correlation matrix $\rho(A,B)$ of two random variables $A$ and $B$, defined as the ratio of the corresponding covariance matrix (cov), divided by the corresponding standard deviations $\sigma_{A,B}$, explicitly given by
\begin{equation}
\begin{split}
    \rho(A,B) &\equiv \dfrac{\text{cov}(A,B)}{\sigma_{A} \sigma_{B}}; 
\\    
    \text{cov}(A,B) &\equiv  \dfrac{1}{N} \sum\limits_{i=1}^{N} (A_{i} - \mu_{A})(B_{i} - \mu_{B}).
\end{split}
\end{equation}
Here, cov reflects the amount of correlation between the magnitudes and directionality (i.e. positive or negative) of random variables deviation from the corresponding mean values $\mu_{A,B}$, normalized by standard deviations $\sigma_{A,B}$. In our case $A$ and $B$ designate several tens of SERS/Raman spectra which belong to group (1) and group (2), respectively, and $N=1021$ is the size of each spectra.} 

\section*{Discussion}

In this work we experimentally and numerically studied the effect of spatial arrangements of DNA bases in ssDNA 3-mers on SERS and Raman spectra. 
Our experimental results, where CE and EM effects were gradually introduced into the system, clearly indicate a difference between the acquired SERS and Raman spectra of sequences which differ under permutation of the DNA bases, and our DFT simulations provide qualitative evidence for dependence of normal mode frequency of DNA base on position along the sequence, with and without CE effect. 
In particular, DFT simulations without metal indicate that frequency dependence and mode splitting are intrinsic property due to complex structure of the molecule, being composed of basic units with permutation dependent molecular conformations; complex interaction with the metal may intensify this difference as it allows higher variability such as different orientation relative to the metal surface. 
\textcolor{black}{Furthermore, we expect that CE effect should play much more significant role in highlighting differences between different permuted molecules than EM effect; the plasmonic mode, which is responsible for the EM enhancement factor, isn't expected to be affected by refractive index changes due to sub-nanometer molecular permutations.}
Nevertheless, more realistic simulations are needed to provide quantitative agreement by incorporating additional effects such as CE due to resonant charge transfer between the adsorbed molecule and the metal substrate, as well as additional dielectric layer between the molecule and the metal. 
While in this work we considered only 3-base long sequences to simplify the system and potentially enhance the effect, we expect this work to facilitate future theoretical studies, as well as experimental TERS-based sensing platforms, which admit sufficiently high spatial resolution to enable excitation of just few bases \cite{downes2006, trautmann2017, liu2019, he2018} to study sensing of longer ssDNA molecules where permuted nucleobases are embedded in a longer DNA sequences. In particular, studying SERS/Raman spectra as a function of intermolecular \cite{zhang2020} or nearest neighbor interaction may potentially enable more accurate SERS-based sensing platforms and allow lower read-out error needed to meet bio-applications \cite{pyrak2019}, support DNA memory applications \cite{panda2018} and allow novel tool to control Raman/SERS response of complex molecules. 
\textcolor{black}{Furthermore, our DFT simulation, presenting mode splitting and normal modes dependence on
the position along the phosphate backbone,} may lead to \textcolor{black}{a novel structure dependent mechanism for} inhomogeneous broadening of the peaks \textcolor{black}{in complex molecules such as ssDNA sequences}, which may be of importance from both fundamental and application perspectives. 

%and considering longer ssDNA molecules. 
% Future works performed in this work both considered simulations of short molecules, requiring super-computer capabilities,
%
%useful for molecular spectroscopy...
%
%useful for DNA as memory application ...
%
%additional interaction effect reported here can improve the sensitivity by lowering the detection error needed to meet clinical requirements \cite{pyrak2019}.

\section*{Funding}

This work was supported by the Defense Advanced Research Projects Agency (DARPA) DSO, NLM, and NAC Programs, the Office of Naval Research (ONR), the National Science Foundation (NSF) grants CBET-1704085, DMR- 1707641, NSF ECCS-180789, NSF ECCS-190184, NSF ECCS-2023730, the Army Research Office (ARO), the San Diego Nanotechnology Infrastructure (SDNI) supported by the NSF National Nanotechnology Coordinated Infrastructure (grant ECCS-2025752), and the ASML-Cymer Corporation.

\textbf{Funding:} This work was supported in part by the National Science Foundation (NSF), the Office of
Naval Research (ONR), the Semiconductor Research Corporation (SRC), the Army Research Office (ARO), DARPA, Cymer Corporation,
and the San Diego Nanotechnology Infrastructure (SDNI) supported by the NSF
National Nanotechnology Coordinated Infrastructure (grant ECCS-1542148).

%\textbf{Author contributions:} S.R., B.H. and Y.F. conceived the project. S.R and B.H. made the research and wrote the manuscript.
%All authors revised the manuscript and contributed to the content.

\textbf{Competing interests:} The authors declare no competing financial interests.

%\textbf{Author contributions:} S.R. and Y.F. conceived the project. S.R. conducted the research and wrote the manuscript. Both authors contributed to the content.

\textbf{Acknowledgments:} This work used the Extreme Science and Engineering Discovery Environment (XSEDE), which is supported by National Science Foundation grant number ACI-1548562.

%%%%%%%%%% Merge with supplemental materials %%%%%%%%%%
% \pagebreak
\widetext
\begin{center}
\newpage
\title{SI}
\textbf{Supplementary Material for: \\ ``The effect of DNA bases permutation on surface enhanced Raman scattering spectrum''}

\text{Shimon Rubin$^{*}$, Phuong H.L. Nguyen$^{*}$, Yeshaiahu Fainman} \\
\text{$^{*}$ Equal contribution} \\
\text{Department of Electrical and Computer Engineering, University of California, San Diego, 9500 Gilman Dr., La Jolla, California 92023, USA} 

\end{center}
%%%%%%%%%% Merge with supplemental materials %%%%%%%%%%

\setcounter{equation}{0}
\setcounter{figure}{0}
\setcounter{section}{0}
\setcounter{table}{0}
\setcounter{page}{1}

\renewcommand{\thesection}{S.\arabic{section}}
\renewcommand{\thesubsection}{\thesection.\arabic{subsection}}
\renewcommand{\thetable}{\arabic{table}} %%%
\makeatletter 
\def\tagform@#1{\maketag@@@{(S\ignorespaces#1\unskip\@@italiccorr)}}
\makeatother
\makeatletter
\makeatletter \renewcommand{\fnum@figure}
{\figurename~S\thefigure}
\makeatother
\makeatletter \renewcommand{\fnum@table}
{\tablename~S\thetable}
\makeatother
%\makeatletter \renewcommand{\fnum@equation}
%{\equationname~S\theequation}
%\makeatother

\section{Dominant modes of AAC and ACC molecules}

DNA molecules are inherently chiral and therefore in addition to ACA, CAA, and CAC, CCA sequences we also consider vibrational modes of AAC and ACC, which are expected to introduce changes in the Raman spectra. Table.S\ref{TableAAC} and Table.S\ref{TableACC} below, describe Raman modes in the spectral regions corresponding to regions specified in Table.1 and Table.2, respectively, with less modes presented in the first spectral region.

\begin{table}[h]
\begin{tabular}{ccccc}

\\[-0.5 ex] 
\textcolor{white}{Name}&  
\begin{tabular}{@{}c@{}}  Frequency [cm$^{-1}$]   \end{tabular} & 
\begin{tabular}{@{}c@{}}  Raman Activity [\r{A}$^{4}$/amu]  \end{tabular} & 
\begin{tabular}{@{}c@{}}  Mode location   \end{tabular} 
\\[1 ex] 

\hline
AAC&  
\begin{tabular}{@{}c@{}} $767.87; 775.38$ \quad \\ $1241.67$ \quad \\ $1376.36; 1376.98$ \quad \\ $1541.47; 1541.96; 1576.51$ \end{tabular} & 
\begin{tabular}{@{}c@{}} $19.85; 55.72$ \quad \\ $21.7$ \quad \\ $48.36; 47.71$ \quad \\ $24.32; 36.80; 37.31; $ \end{tabular} & 
\begin{tabular}{@{}c@{}} $3^{\prime}$ sugar, $3^{\prime}$PHB, C; C \quad \\ C, $3^{\prime}$sugar \quad \\ cen A, cen sugar; $5^{\prime}$sugar \quad \\ $5^{\prime}$A; cen A; C \end{tabular} 
\\ [1ex]

\hline
AAC-Ag$_{12}$&  
\begin{tabular}{@{}c@{}}  $784.36; 772.09$ \quad \\ $1247.22$ \quad \\  $1370.59; 1372.23$ \quad \\ $1542.26; 1545.89; 1570.77$ \end{tabular} & 
\begin{tabular}{@{}c@{}}  $83.76; 30.48$ \quad \\ $16.40$ \quad \\ $106.15; 108.84$ \quad \\ $115.58; 106.19; 191.45 $ \end{tabular} & 
\begin{tabular}{@{}c@{}}  C; $3^{\prime}$sugar, $3^{\prime}$PHB, cen sugar \quad \\ C, $3^{\prime}$sugar \quad \\ $5^{\prime}$A, $5^{\prime}$ sugar; cen A, cen sugar \quad \\ $5^{\prime}$A; cen A; C \end{tabular} 
\\ [1ex]

\\ [-2.5ex]
\hline
\end{tabular}    
\caption{DFT simulation results presenting frequencies, RAs and the corresponding Raman active molecular groups in AAC and AAC-Ag$_{12}$ molecules which belong to group (1).}
\label{TableAAC}
\end{table}

\begin{table}[h]
\begin{tabular}{ccccc}

\\[-0.5 ex] 
\textcolor{white}{Name}&  
\begin{tabular}{@{}c@{}}  Frequency [cm$^{-1}$]   \end{tabular} & 
\begin{tabular}{@{}c@{}}  Raman Activity [\r{A}$^{4}$/amu]  \end{tabular} & 
\begin{tabular}{@{}c@{}}  Mode location   \end{tabular} 
\\[1 ex] 

\hline
ACC&  
\begin{tabular}{@{}c@{}}  $776.46; 779.19$ \quad \\ $1230.42; 1242.28$ \quad \\ $1376.59$ \quad \\ $1542.21; 1573.07; 1576.45$  \end{tabular} & 
\begin{tabular}{@{}c@{}}  $36.05; 22.86$ \quad \\ $47.48; 19.97$ \quad \\ $51.25$ \quad \\ $28.43; 33.87; 34.46$   \end{tabular} & 
\begin{tabular}{@{}c@{}}  $3^{\prime}$C; cen C, $3^{\prime}$C \quad \\ cen C, cen sugar; $3^{\prime}$C, 3$^{\prime}$ sugar \quad \\ A, $5^{\prime}$ sugar \quad \\ A; cen C; $3^{\prime}$C  \end{tabular} 
\\ [1ex]

\hline
%\\
%\\[-6.5ex] 
ACC-Ag$_{12}$& 
\begin{tabular}{@{}c@{}} $788.32; 791.96$ \quad \\ $1232.33$ \quad \\ $1387.99$ \quad \\ $1561.19; 1570.39; 1572.39$ \end{tabular} & 
\begin{tabular}{@{}c@{}} $70.31; 47.72$ \quad \\ $22.95$ \quad \\ $21.99$ \quad \\ $65.88; 90.51; 48.26$ \end{tabular} & 
\begin{tabular}{@{}c@{}} cen C; 3$^{\prime}$C  \quad \\ cen C, cen sugar \quad \\ A \quad \\ A; cen C; 3$^{\prime}$C \end{tabular}
\\ 
\\ [-2.5ex]
\hline
\end{tabular}    
\caption{DFT simulation results presenting frequencies, RAs and the corresponding Raman active molecular groups in ACC and ACC-Ag$_{12}$ molecules which belong to group (2).}
\label{TableACC}
\end{table}

Fig.S\ref{TableAAC} presents the effect of partial bonding of Ag$_{4}$ nanoparticles on the emerging SERS spectra.

\clearpage

\section{Spatial Conformations of ACA, CAA, CAC, and CCA molecules}

\textcolor{black}{The following image presents computational results of the optimized geometry for group (1) and group (2) ssDNA molecules.}

\begin{figure}[h]
	\includegraphics[scale=0.09]{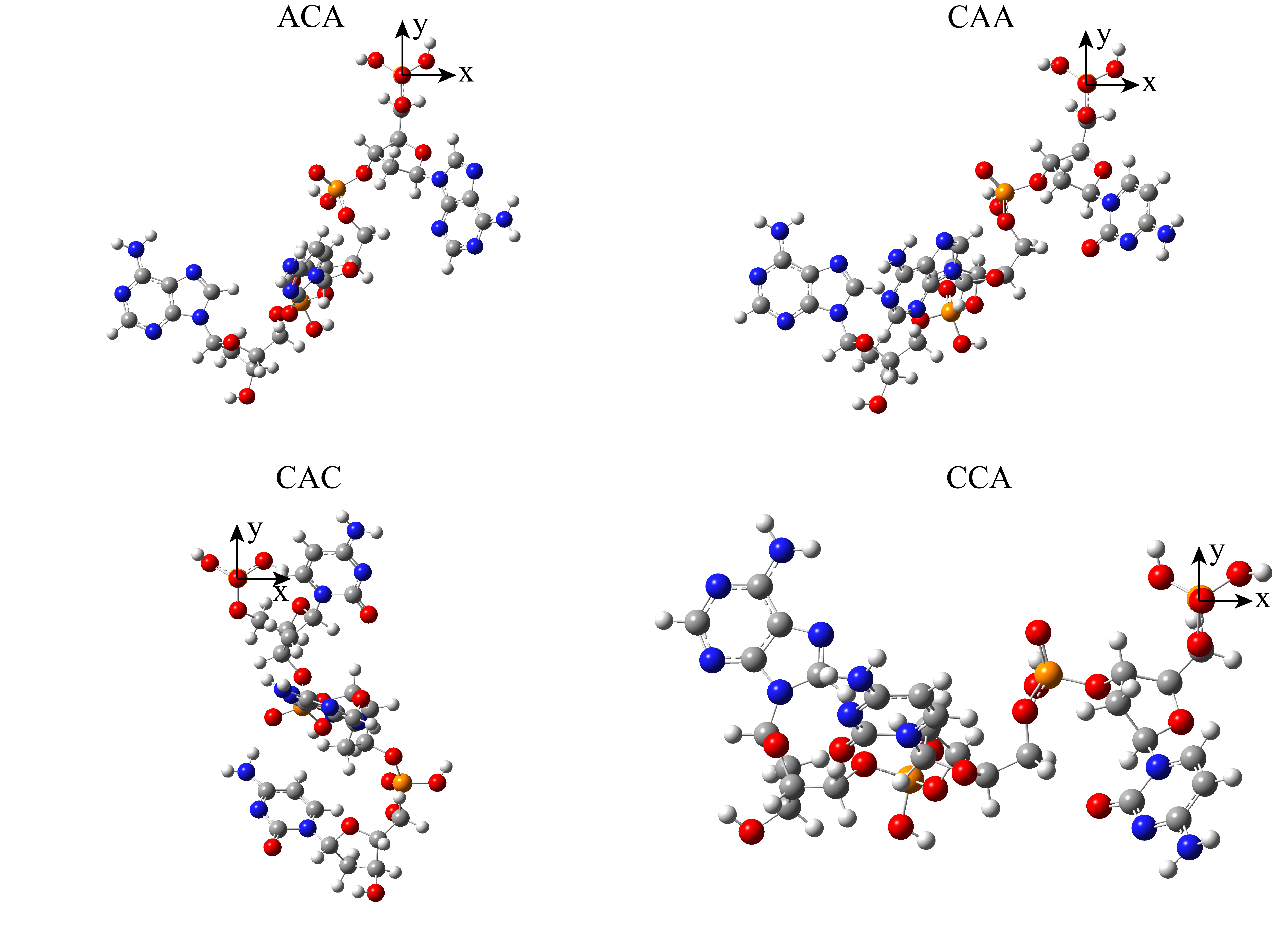}
    \caption{\textcolor{black}{DFT results presenting optimized geometry of group (1) and group (2) ssDNA molecules in the first and second rows, respectively. Red - oxygen (O), blue - nitrogen (N), grey - carbon (C), white - hydrogen (H), orange - phosphorous (P). In all molecules the orientation (coordinate system) was fixed by fixing the orientation of the four O's which are bond to P at $5^{\prime}$ end. In particular P is chosen as origin; x-axis is parallel to a pair of O's where each O is connected to a single P and H atoms; y-axis is antiparallel to the P-O bond where O is bond to C; z-axis (towards the reader) is along the P=0 bond.}}
    \label{Conformation}
\end{figure}

\section{Additional details about fabricated nanorod array}

\begin{figure}[h]
	\includegraphics[scale=0.5]{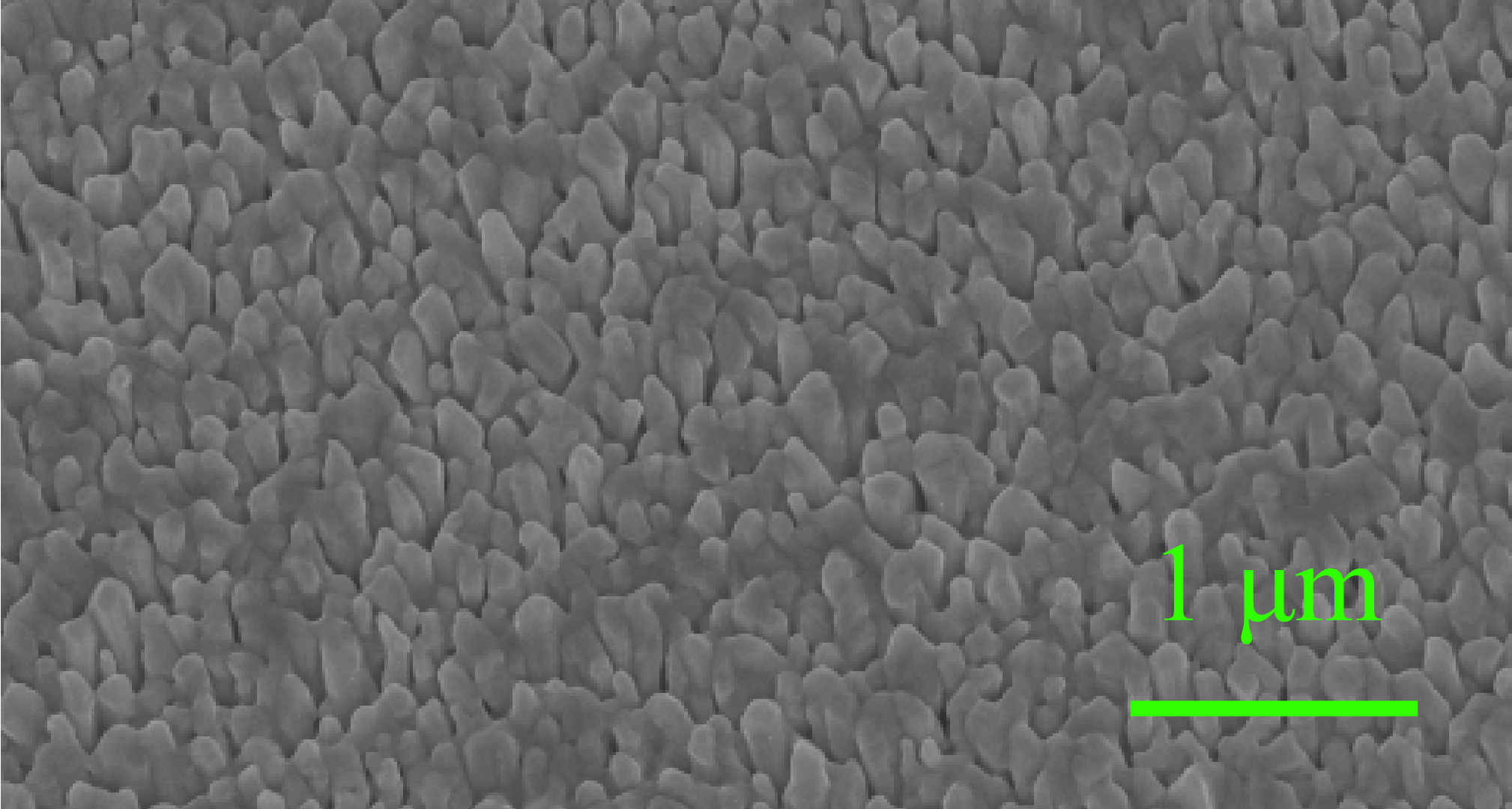}
    \caption{\textcolor{black}{SEM image showing typical morphology of the OAD-fabricated gold nanorod array.}}
    \label{SEMgold}
\end{figure}

\textcolor{black}{Fig. S\ref{SEMgold} presents SEM image of a typical OAD-fabricated nanorod array. The average height of the nanorods, is approximately $h = 200$ nm, whereas the average diameter of the rods is approximately $\Lambda = 50$ nm for gold (and $\Lambda = 100$ nm for silver)}

\textcolor{black}{Based on the width of the error bars, which represent standard deviation due to measurements acquired at 25 different spatial locations on the substrate, relative to the mean peak height, we conclude that the mean signal variation across all spectral domains is approximately 6\%.}

\textcolor{black}{Based on our recent work [41], where identical nanorod arrays were employed, we bring here the corresponding values of the CE factors. In particular, these are approximately $\sim 2-20$ for gold (depending on the spectral region) and $\sim 2$ for silver. The latter were obtained by comparing between SERS spectra of molecules adsorbed directly to the metal nanorod array to similar array coated with a $2$ nm thick Al$_{2}$O$_{3}$ film, which eliminates potential resonant charge transfer effect. Experimental measurement of total enhancement factor, due to both EM and CE effects, requires determination of surface concentration of adsorbed ssDNA molecules to the nanorod array after the rinsing step, which is beyond the scope of this work.}

\section{Estimation for the EM enhancement factor}

We define the EM enhancement factor (EF) for each one of the 3-mers employed in this work as
\begin{equation}
    EF = \bar{I}_{SERS}/\bar{I}_{Raman},
\end{equation}
where $\bar{I}_{SERS}$ and $\bar{I}_{Raman}$ represent "signal per molecule", i.e. the integral of SERS (or Raman) spectra ($I_{SERS}, I_{Raman}$) over the relevant spectral domain divided by the corresponding number of molecules ($N_{SERS}, N_{Raman}$) in the illuminated region and also by the illumination power $(P_{SERS}, P_{Raman})$, allowing to write the equation above as
\begin{equation}
    EF = \dfrac{I_{SERS}}{N_{SERS} \cdot P_{SERS}} \bigg/ \dfrac{I_{Raman}}{N_{Raman} \cdot P_{Raman}}.
\end{equation}
Based on our experimental spectra, the ratios of the signals $I_{SERS}/I_{Raman}$ for the sequences ACA, CAA, CAC, CCA are: $28.2$, $30.9$, $22.0$, $42.2$, respectively; i.e. very similar and differ at most by a factor approximately two. Next, $P_{SERS} = 10$ mW, $P_{Raman} = 100$ mW, leading to $P_{SERS}/P_{Raman}=10$. Finally, we determine the number of active molecules in each one of the cases. Assuming the 3-base molecules cover uniformly the $5 \times 5 = 25$ $\mu$m$^{2}$ sampling area, and the in-plane area of the molecule is 
\begin{equation}
    \text{length} \times \text{width} = (0.676 \text{ nm} \times 3 \text{ bases}) \times 0.9 \text{ nm} = 1.82 \cdot 10^{-6} \mu \text{m}^{2}, 
\end{equation}
yields the following number of molecules in the sampling area, $N_{SERS} = 25/(1.82 \cdot 10^{-6}) = 1.37 \cdot 10^{7}$. To determine $N_{Raman}$, defined by,
\begin{equation}
    N_{Raman} = \text{illumination volume} \times \text{copies per volume},
\end{equation}
we estimate illumination volume as $\pi \omega_{0}^{2} L_{R} = 2.36 \cdot 10^{-13}$ l. Here, $\omega_{0} = \lambda/(\pi \cdot NA) = 785 \text{ nm}/(\pi \cdot 0.12) = 2082.3$ nm  is the radius of the Gaussian beam waist, $\lambda$ is the wavelength of illumination laser in the free space, and NA is the corresponding numerical aperture; $L_{R} = \pi \omega_{0}^{2}/\lambda = 17352.7$ nm is the Rayleigh length. The number of molecules per volume is a product of molar concentration, $100$ \text{$\mu$}M $=10^{-4}$ m/l, with Avogadro number, $N_{A} = 6.02 \cdot 10^{23}$ copies/m, leading to $N_{Raman} = 1.42 \cdot 10^{7}$ copies. Inserting, these parameters into the definition of the EF, yields the following similar values,
\begin{equation}
    EF = 2.93 \cdot 10^{2}, 3.20 \cdot 10^{2}, 1.04 \cdot 10^{2}, 4.37 \cdot 10^{2} 
\end{equation}
for the sequences ACA, CAA, CAC, CCA, respectively.
\end{document}